\documentclass{article}
\usepackage{spconf,amsmath,amssymb,graphicx,booktabs}
\usepackage[hidelinks]{hyperref}

\title{ReaFlow-TTS: Realization-Conditioned Flow Matching for High-Quality and Controllable Speech Synthesis}
\name{Junyi Zhao, Changsheng Ma, Yihao Qin, Yongfeng Tao, Minqiang Yang$^{*}$, Hu Bin$^{*}$\thanks{$^{*}$Corresponding authors.}}

\address{School of Information Science and Technology, Lanzhou University, China}
\begin{document}
\raggedbottom
 \ninept
\maketitle
\begin{abstract}

In flow-matching text-to-speech (TTS), different speech
realizations can induce different target velocities under the
same generation conditions.
A deterministic velocity field trained with squared error
predicts their conditional mean, thereby marginalizing
realization-dependent variation.
Meanwhile, modeling such variation does not inherently
provide a semantically interpretable interface for attribute
manipulation.
We propose ReaFlow-TTS, a realization-conditioned
flow-matching framework that 
introduces an utterance-level stochastic realization latent and uses it to condition velocity prediction throughout the generation trajectory.
We further impose valence--arousal--dominance (VAD) semantics
on the realization space, enabling direct and graded
attribute manipulation without target speech at inference.
Experiments demonstrate improved synthesis quality over a
matched full-mask baseline and reproducible latent-induced
pitch, energy, and timing tendencies across initial-noise
samples, providing behavioral evidence that the latent is
used as a reusable realization condition.
Subjective evaluation further demonstrates graded VAD
manipulation across generation contexts with only modest changes in naturalness.
\end{abstract}
\begin{keywords}
Text-to-Speech, Flow Matching, Speech Realization,
Expressive Speech Synthesis
\end{keywords}
\section{Introduction}
\label{sec:introduction}

Neural TTS systems have achieved high
intelligibility and naturalness~\cite{shen2018natural,
kim2021conditional,wang2025maskgct,chen2025f5}, with increasing
attention to expressive and controllable speech
generation~\cite{ren2020fastspeech,fujita2025voice,
rautenberg2025speech}.
The same text can be rendered with different timbres,
prosodic patterns, and emotions. We refer to each such
rendering as a \emph{speech realization}.
Existing methods capture such variation through reference
speech, prosodic representations~\cite{kim2021conditional,
ju2024naturalspeech,wang2025prosodyflow,
nguyen2025diflow}, internal feature
guidance~\cite{xie2025emosteer,li2026restyle}, and
velocity-field design~\cite{lee2025vector,pankov2026pfluxtts}.

\begin{figure}[t]
    \centering
    \includegraphics[width=0.9\columnwidth]{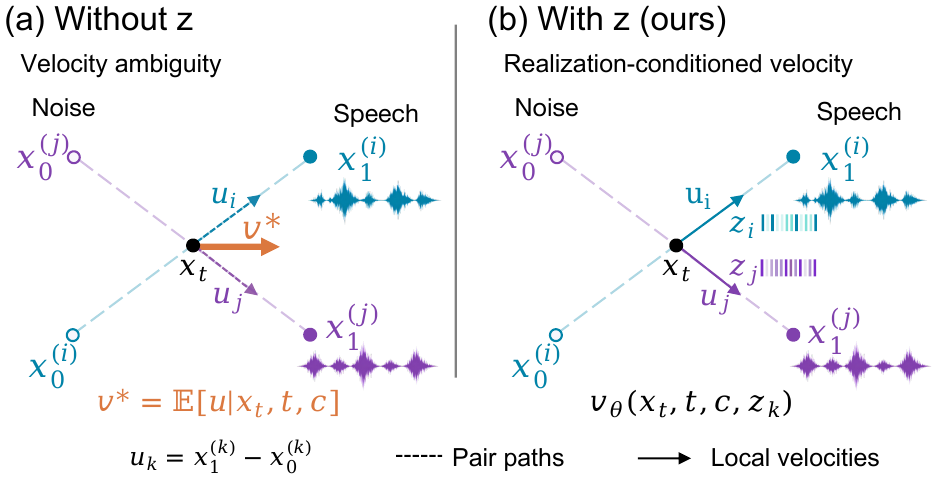}
    \caption{Illustration of realization-dependent velocity modeling in flow-matching TTS. (a) Different speech realizations can induce different target velocities under the same generation conditions, which are conditionally averaged by a deterministic velocity predictor. (b) ReaFlow-TTS introduces a realization latent $z$ to explicitly condition velocity prediction on realization information.}
    \label{fig:intro}
\end{figure}

In flow matching, different speech realizations can induce
variation in the target velocity field.
Variational Rectified Flow Matching (VRFM) \cite{ref18} shows that paths induced by different source--target pairs may intersect at the same state and flow time while having different target velocities, giving rise to velocity ambiguity.
As illustrated in Fig.~\ref{fig:intro}(a), the same phenomenon
can arise across paths associated with different speech
realizations, yielding different target velocities under the
same state, flow time, and generation conditions.
Under a squared-error objective, a deterministic velocity
network predicts their conditional mean, marginalizing
realization-dependent variation rather than explicitly
distinguishing the corresponding generation directions.

Explicitly modeling realization variation, however, raises a further question: whether the resulting representation can also support interpretable manipulation. A speech realization jointly reflects multiple acoustic attributes, yet variation in its representation does not necessarily align with semantically interpretable attributes. Consequently, even when realization variation is explicitly modeled, manipulating a specific attribute in a direct and graded manner remains non-trivial~\cite{fujita2025voice,li2026restyle}. This motivates a realization representation that can both condition velocity prediction and serve as a semantic interface for graded attribute manipulation.


To this end, we propose ReaFlow-TTS, \textbf{Rea}lization-Conditioned \textbf{Flow} Matching for \textbf{TTS}, which
represents speech realization with an utterance-level
stochastic latent variable $z$ and explicitly conditions the
velocity field on this realization information.
The same $z$ is maintained throughout generation, providing
a consistent realization condition across the flow trajectory.
During training, a posterior encoder infers $z$ from the
target Mel spectrogram and regularizes its distribution toward a standard Gaussian prior, allowing direct prior sampling at inference.
We further learn a linear semantic mapping $W$ that structures the realization space with VAD semantics, enabling direct and graded attribute manipulation~\cite{liu2025uddetts,cho2025emosphere++}.


Experiments demonstrate improved intelligibility and naturalness, together with reproducible latent-induced acoustic tendencies across initial-noise samples and graded VAD manipulation across generation contexts. The main contributions of this work are as follows: 1) We introduce an utterance-level stochastic realization latent as a trajectory-level condition for flow-matching velocity prediction, allowing the model to explicitly account for realization-dependent variation while supporting direct prior sampling at inference. 2) We semantically structure the same realization space with VAD attributes, providing a lightweight interface for direct and graded attribute manipulation across generation contexts.

\section{PRELIMINARIES}
\label{sec:format}

Let $p_0$ and $p_1$ denote the source and target distributions,
respectively.
Rectified Flow Matching~\cite{lipman2022flow,liu2022flow}
learns a velocity field $v_\theta(x_t,t)$ that transports
samples from $p_0$ to $p_1$ over $t\in[0,1]$.
Given independently sampled $x_0\sim p_0$, $x_1\sim p_1$,
and $t\sim\mathcal{U}(0,1)$, the linear interpolation path
and its target velocity are
\begin{equation}
\begin{gathered}
x_t = \phi(x_0,x_1,t) = (1-t)x_0+t x_1,\\[2pt]
v(x_0,x_1,t)
= \frac{\partial\phi(x_0,x_1,t)}{\partial t}
= x_1-x_0.
\end{gathered}
\label{eq:rf_path}
\end{equation}
The velocity field is learned by minimizing
\begin{equation}
\mathcal{L}_{\mathrm{FM}}
=
\mathbb{E}_{t,x_0,x_1}
\left[
\left\|
v_\theta(x_t,t)-v(x_0,x_1,t)
\right\|_2^2
\right].
\label{eq:fm_loss}
\end{equation}

Different source--target pairs can induce different target
velocities at the same state and flow time.
Under the squared-error objective, the optimal deterministic
velocity predictor is therefore the conditional mean
\begin{equation}
v^*(x_t,t)
=
\mathbb{E}
\left[
v(x_0,x_1,t)\mid x_t,t
\right].
\label{eq:optimal_velocity}
\end{equation}
Consequently, a deterministic velocity field represents the
conditional average of the target velocities rather than
explicitly modeling their underlying variation.

\section{Method}
\label{sec:method}

\subsection{Overview}
\label{sec:overview}

Fig.~\ref{fig:main1} illustrates ReaFlow-TTS, which uses an
utterance-level realization latent $z$ for two coupled roles.
First, $z$ conditions the velocity field on realization
information and provides a consistent realization condition
throughout generation. Second, the same latent space is semantically structured with VAD attributes, enabling graded attribute manipulation.
During training, $z$ is inferred from target speech and regularized toward a standard Gaussian prior, supporting direct prior sampling at inference.

\subsection{Realization-Conditioned Velocity Modeling}
\label{sec:velocity_modeling}

\begin{figure}[t]
    \centering
    \includegraphics[width=0.9\columnwidth]{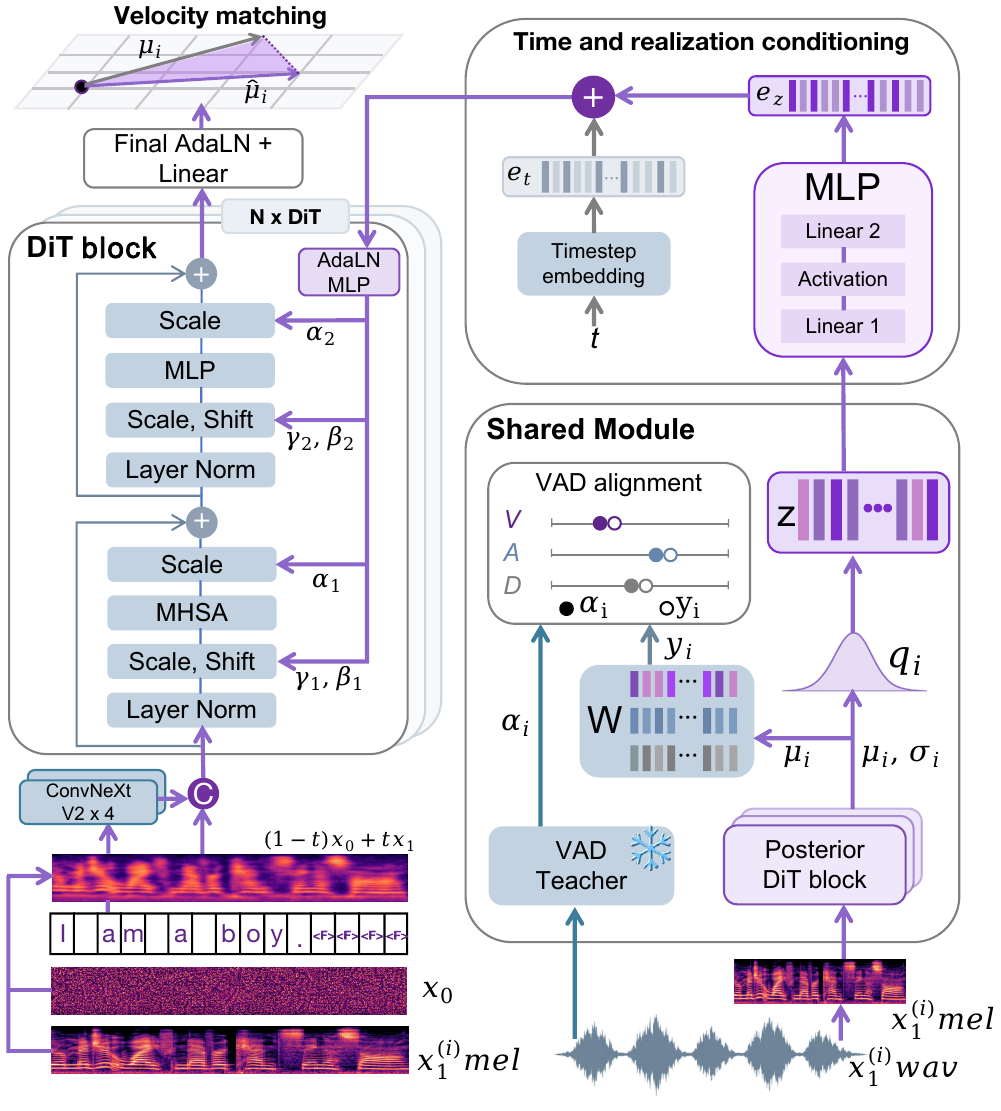}
    \caption{Overview of ReaFlow-TTS. The realization latent $z$ conditions velocity prediction throughout the flow trajectory, while VAD supervision structures the latent space for graded attribute manipulation. $C$ and $+$ denote concatenation and addition, respectively.}
    \label{fig:main1}
\end{figure}

In TTS, let $x_1\in\mathbb{R}^{L\times F}$ denote the target
Mel spectrogram with $L$ acoustic frames and $F$ Mel-frequency
bins, and let $c$ denote the generation conditions, including
text and length.
Given $x_0\sim p_0$ and $t\sim\mathcal{U}(0,1)$, the flow
state $x_t$ and target velocity $v=x_1-x_0$ follow
Eq.~\eqref{eq:rf_path}.
Extending the conditional-mean property in
Sec.~\ref{sec:format} to conditional flow matching, a
deterministic velocity predictor estimates
$\mathbb{E}[v\mid x_t,t,c]$, marginalizing target-velocity
variation associated with different speech realizations.

To explicitly condition velocity prediction on such variation,
we introduce a $d_z$-dimensional utterance-level realization
latent $z\in\mathbb{R}^{d_z}$ and model the velocity field as
$v_\theta(x_t,t,c,z)$.
During training, $z$ is inferred from the observed realization
$x_1$ using a diagonal-Gaussian posterior
\begin{equation}
q_\psi(z\mid x_1)
=
\mathcal{N}\!\left(
\mu_\psi(x_1),
\operatorname{diag}(\sigma_\psi^2(x_1))
\right),
\end{equation}
with reparameterized sampling~\cite{kingma2014autoencoding}
\begin{equation}
z=\mu_\psi(x_1)+\sigma_\psi(x_1)\odot\epsilon,
\qquad
\epsilon\sim\mathcal{N}(0,I_{d_z}).
\end{equation}

As illustrated in Fig.~\ref{fig:main1}, the sampled $z$ is
projected into a realization embedding and combined with the
timestep embedding to modulate each DiT block~\cite{peebles2023scalable} through AdaLN.
As an utterance-level latent, $z$ is shared across acoustic frames during training and held fixed across flow time during generation, providing a consistent realization condition throughout the generation trajectory.

We regularize the posterior toward a standard Gaussian prior
$p(z)=\mathcal{N}(0,I_{d_z})$ and optimize
\begin{equation}
\begin{aligned}
\mathcal{L}_{\mathrm{base}}
=
\mathbb{E}_{c,t,x_0,x_1}
\Bigg[
&\mathbb{E}_{z\sim q_\psi(z\mid x_1)}
\left[
\left\|
v_\theta(x_t,t,c,z)-v
\right\|_2^2
\right]\\
&+
\beta D_{\mathrm{KL}}\!\left(
q_\psi(z\mid x_1)\,\|\,p(z)
\right)
\Bigg].
\end{aligned}
\label{eq:base_loss}
\end{equation}
The KL term enables realization latents to be sampled directly
from the prior at inference, without access to target speech.

\subsection{Semantic Structuring of the Realization Space}
\label{sec:semantic_structuring}

While $z$ provides realization information for velocity
prediction, its latent space is not inherently aligned with
interpretable speech attributes.
We therefore impose VAD semantics on the realization space
using the shared module in Fig.~\ref{fig:main1}.
For utterance $i$, we use the posterior mean
$\mu_i=\mu_\psi(x_1^{(i)})$ and learn a linear mapping
$W\in\mathbb{R}^{3\times d_z}$ to predict its valence,
arousal, and dominance attributes:
\begin{equation}
y_i = W\mu_i.
\label{eq:vad_readout}
\end{equation}
A frozen VAD teacher extracts the corresponding attribute
scores from the waveform, which are standardized using
training-set statistics to obtain the supervision target
$\alpha_i$.
The posterior mean is used for semantic supervision to avoid
stochastic variation introduced by latent sampling.

To additionally capture relative attribute differences, we
construct pairs $(i,j)\sim\mathcal{P}$ from utterances with
the same text but different speech realizations.
This pairing controls linguistic content when comparing
realization-dependent VAD variation.
Defining
$\Delta y_{ij}=y_j-y_i$ and
$\Delta\alpha_{ij}=\alpha_j-\alpha_i$, we optimize
\begin{equation}
\begin{aligned}
\mathcal{L}_{\mathrm{attr}}
&=
\mathbb{E}_{(i,j)\sim\mathcal{P}}
\left[
\frac{
\operatorname{SmoothL1}(y_i,\alpha_i)
+
\operatorname{SmoothL1}(y_j,\alpha_j)
}{2}
\right],\\
\mathcal{L}_{\mathrm{dist}}
&=
\mathbb{E}_{(i,j)\sim\mathcal{P}}
\left[
\left(
\|\Delta y_{ij}\|_2
-
\|\Delta\alpha_{ij}\|_2
\right)^2
\right].
\end{aligned}
\label{eq:semantic_losses}
\end{equation}
$\mathcal{L}_{\mathrm{attr}}$ establishes absolute VAD
alignment, while $\mathcal{L}_{\mathrm{dist}}$ preserves
relative VAD distances between paired realizations.
Together, they provide a semantically calibrated realization
space for the graded manipulation described in
Sec.~\ref{sec:generation_control}.
The complete training objective is
\begin{equation}
\mathcal{L}_{\mathrm{total}}
=
\mathcal{L}_{\mathrm{base}}
+
\lambda_{\mathrm{attr}}\mathcal{L}_{\mathrm{attr}}
+
\lambda_{\mathrm{dist}}\mathcal{L}_{\mathrm{dist}}.
\label{eq:total_objective}
\end{equation}
The velocity network, posterior encoder, and $W$ are jointly
optimized, while the VAD teacher remains frozen.

\subsection{Realization Sampling and VAD Manipulation}
\label{sec:generation_control}



At inference, the posterior encoder and VAD teacher are
discarded.
A realization latent is sampled directly from the prior,
$z_{\mathrm{base}}\sim\mathcal{N}(0,I_{d_z})$, independently
of the initial noise $x_0$.
The sampled $z_{\mathrm{base}}$ is held fixed throughout ODE
integration, providing a consistent realization condition
across flow time.
Thus, $x_0$ initializes the generation trajectory, while
$z_{\mathrm{base}}$ provides the sampled realization condition for the velocity field.

The semantic mapping $W$ learned in
Sec.~\ref{sec:semantic_structuring} further provides an
interface for manipulating the sampled realization.
Given a desired change in VAD attributes
$\Delta\alpha\in\mathbb{R}^{3}$, we obtain the corresponding
latent adjustment using a regularized right inverse of $W$:
\begin{equation}
\Delta z
=
W^\top
\left(
WW^\top+\rho I_3
\right)^{-1}
\Delta\alpha,
\qquad
z'=z_{\mathrm{base}}+\Delta z,
\label{eq:latent_editing}
\end{equation}
where $\rho$ is a regularization coefficient.
For manipulation along attribute
$k\in\{V,A,D\}$, we set
$\Delta\alpha=\eta e_k$, where $e_k$ denotes the corresponding
unit vector and $\eta$ controls the manipulation strength.
The edited latent $z'$ replaces $z_{\mathrm{base}}$ and remains
fixed throughout the generation trajectory.
Setting $\eta=0$ recovers the base realization.

\section{Experiments}
\label{sec:experiments}

\subsection{Experimental Setup}
\label{sec:experimental_setup}

\subsubsection{Datasets.}
We first trained on the filtered LibriTTS
corpus~\cite{zen2019libritts} (554~h), followed by the
English subset of ESD~\cite{zhou2021seen} for VAD semantic structuring.
The ESD subset contains 17,500 parallel utterances from
10 speakers across five emotions, enabling the same-text pairing in Sec.~\ref{sec:semantic_structuring}.
We held out all 50 utterances corresponding to one text for
latent usage diagnostics and used the remainder for training.
Synthesis quality was evaluated on 1,127 same-speaker,
cross-sentence pairs from F5-TTS~\cite{chen2025f5}
(LibriSpeech-PC test-clean~\cite{meister2023librispeechpc}).

\subsubsection{Comparison models.}
F5-TTS and F5-TTS (full mask) served as the primary baselines, with the latter isolating the effect of full acoustic masking from realization conditioning. ZipVoice base~\cite{zhu2025zipvoice} served as an external reference using its public checkpoint and official zero-shot inference procedure. ReaFlow-TTS (500k) denotes the first-stage checkpoint before VAD supervision.

\subsubsection{Model configuration and training.}

Both F5-TTS baselines and ReaFlow-TTS used 24-kHz audio, 100-dimensional log-Mel spectrograms, and comparable 158M-parameter inference models. F5-TTS retained its original acoustic masking strategy, whereas F5-TTS (full mask) and ReaFlow-TTS used full acoustic masking. ReaFlow-TTS used a 256-dimensional realization latent and a 9-block DiT posterior encoder, which was discarded at inference.

All models used AdamW with a learning rate of $7.5\times10^{-5}$ and were trained for 500k steps on LibriTTS followed by 100k steps on ESD. ReaFlow-TTS used $\mathcal{L}_{\mathrm{base}}$ with
$\beta=3\times10^{-4}$ in the first stage and
$\mathcal{L}_{\mathrm{total}}$ with $\beta=10^{-3}$
and $\lambda_{\mathrm{attr}}=\lambda_{\mathrm{dist}}=0.05$
in the second. The F5-TTS baselines used the standard flow-matching loss. VAD supervision used a frozen audEERING Wav2Vec2 model~\cite{wagner2023dawn} fine-tuned on MSP-Podcast~\cite{lotfian2017building}.

\subsubsection{Generation settings.}
The quality comparison used identical reference--target pairs
without access to the target recordings or durations.
F5-TTS used the reference Mel spectrogram as an acoustic
prefix, whereas F5-TTS (full mask) and ReaFlow-TTS fully
masked the reference acoustic condition.
Both F5-TTS baselines sampled the initial noise $x_0$;
ReaFlow-TTS independently sampled $x_0$ and
$z\sim\mathcal{N}(0,I_{d_z})$, with $z$ fixed throughout
ODE integration.
All models used the same length estimation, 32-step Euler
sampling (CFG${}=2.0$), and Vocos
vocoder~\cite{siuzdak2024vocos}.
For VAD manipulation, we set the regularization coefficient
in Eq.~\eqref{eq:latent_editing} to $\rho=10^{-3}$.

\subsubsection{Evaluation metrics.}
For objective quality evaluation, three samples were generated per condition.
WER computed using Whisper-large-v3~\cite{radford2023robust}
and UTMOS~\cite{saeki2022utmos} measured content accuracy and
predicted naturalness, respectively.
Thirty proficient English listeners completed subjective
evaluations on the same 20 unseen texts.
Five-point NMOS assessed naturalness, while seven-point ratings
assessed perceived VAD attribute intensities.
Higher scores indicated greater naturalness and stronger
perceived attribute intensity, respectively.

For the subjective evaluation of ReaFlow-TTS, three base
realization latents were shared across the 20 texts, with two
initial noise samples per text, yielding 120 unedited
utterances.
These utterances were used for both NMOS evaluation and the
zero-strength VAD condition, with naturalness ratings reused.
VAD manipulation was additionally evaluated by attribute-ranking
accuracy using pairwise forced-choice judgments.
Each utterance received 10 valid ratings.

\begin{figure*}[t] 
\centering
\includegraphics[width=0.9\textwidth]{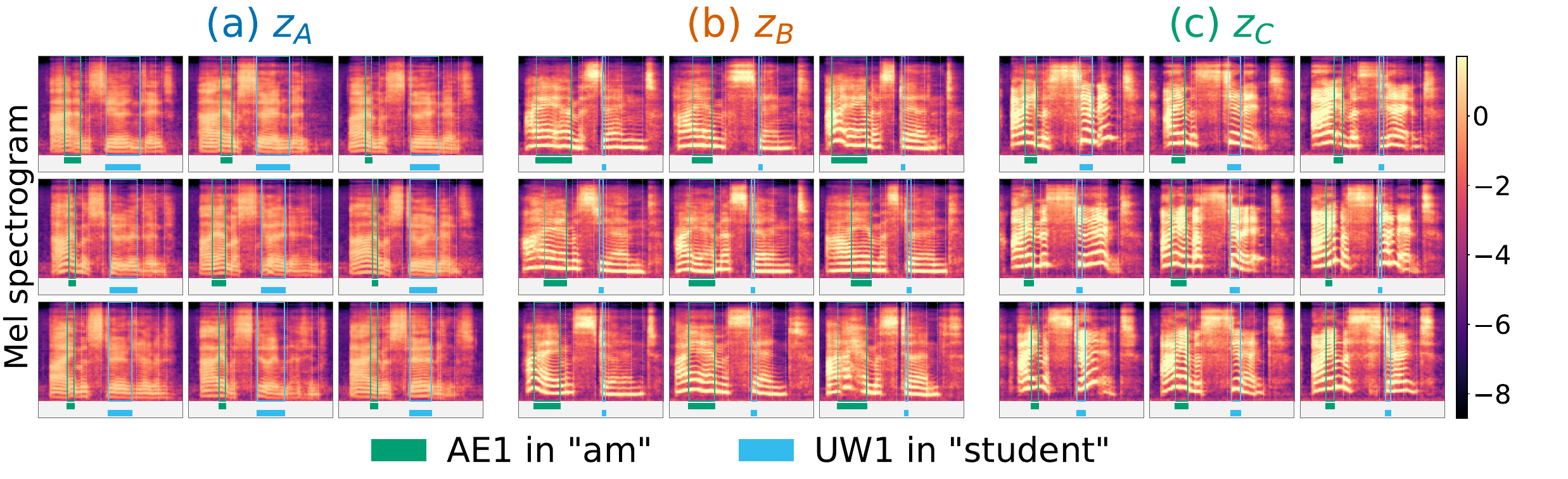} 
\caption{Cross-noise comparison of three realization latents across nine matched initial-noise samples. Corresponding grid positions across the three latent conditions share the same $x_0$. Shared log-Mel scale: approximately $-8$ to $0$.}
\label{fig:cross_noise_reuse}
\end{figure*} 

\subsection{Speech Quality and VAD Manipulation}
\label{sec:quality_control}


\begin{table}[t]
\centering
\caption{Speech quality and inference parameter counts.}
\label{tab:speech_quality}
\small
\setlength{\tabcolsep}{2pt}
\renewcommand{\arraystretch}{1.18}
\begin{tabular}{@{}lcccc@{}}
\toprule
System
& \shortstack{Inf.\\params.}
& WER (\%) $\downarrow$
& UTMOS $\uparrow$
& NMOS $\uparrow$ \\
\midrule
Ground truth
& --- & 2.28 & 4.10 & 3.93 \\
F5-TTS
& 157.97M & 2.27 & 3.91 & 3.72 \\
F5-TTS (full mask)
& 157.97M & 2.32 & 3.83 & 3.67 \\
ZipVoice base
& 122.66M & 2.24 & 3.87 & 3.63 \\
\addlinespace[2pt]
ReaFlow-TTS (500k)
& 158.75M & 2.08 & 4.19 & 3.93 \\
ReaFlow-TTS (full)
& 158.75M & 2.09 & 4.19 & 3.91 \\
\bottomrule
\end{tabular}
\end{table}

\subsubsection{Speech quality.}
Table~\ref{tab:speech_quality} compares synthesis quality
without VAD manipulation.
Full acoustic masking degrades all three quality metrics for
F5-TTS, indicating that removing visible reference acoustics
makes generation more challenging.
Under the same full-mask setting, ReaFlow-TTS not only
recovers this degradation but also surpasses the original
F5-TTS.
This comparison suggests that the quality improvement is
associated with realization-conditioned velocity modeling
rather than full acoustic masking alone.
Moreover, ReaFlow-TTS (500k) already exhibits nearly the same quality as the full model, showing that the observed quality improvement is already present before semantic calibration and is largely preserved after VAD supervision.

\subsubsection{Graded VAD manipulation.}
We evaluate VAD manipulation on the 120 base contexts
described in Sec.~\ref{sec:experimental_setup}.
Within each context, the text, generation length, $x_0$, and
$z_{\mathrm{base}}$ are fixed, while
$\Delta\alpha=\eta e_k$ is varied with
$\eta\in\{0,0.5,1\}$ along each VAD axis.
As shown in Table~\ref{tab:vad_control}, arousal and dominance
exhibit clear graded perceptual responses as $\eta$ increases,
showing that the magnitude of latent displacement translates
into corresponding changes in perceived attribute intensity.
Valence is less sensitive to the smaller displacement but
shows clearer separation at the larger strength, suggesting
unequal sensitivity across the learned VAD directions.
Pairwise ordering results further support the reuse of the learned VAD directions across different texts, base realization latents, and initial noise samples. Meanwhile, NMOS remains relatively stable across manipulation strengths, with only modest changes in naturalness.


\begin{table}[t]
\centering
\caption{Perceptual attribute changes and naturalness under graded VAD manipulation.}
\label{tab:vad_control}
\setlength{\tabcolsep}{2pt}
\renewcommand{\arraystretch}{1.08}
\begin{tabular*}{\columnwidth}{@{\extracolsep{\fill}}lccc@{}}
\toprule
Axis
& \shortstack{Attribute\\score\\[-1pt]
\scriptsize $\eta=0,\,0.5,\,1$}
& \shortstack{NMOS $\uparrow$\\[-1pt]
\scriptsize $\eta=0,\,0.5,\,1$}
& \shortstack{Ordering\\acc. (\%) $\uparrow$\\[-1pt]
\scriptsize $(0,0.5),\ (0.5,1),\ (0,1)$} \\
\midrule
V
& 3.62/3.77/4.21
& 3.91/3.93/3.83
& 61/78/82 \\
A
& 3.38/4.76/5.23
& 3.91/3.82/3.80
& 89/81/92 \\
D
& 3.71/4.98/5.36
& 3.91/3.84/3.77
& 86/82/88 \\
\bottomrule
\end{tabular*}
\end{table}


\subsection{Latent Usage and Cross-Noise Reusability}
\label{sec:Latent Usage and Cross-Noise Reusability}


To examine whether $z$ serves as a reusable realization condition rather than producing effects specific to a particular initial-noise trajectory, we fix the synthesis text to ``I am a student.'' and cross three realization latents ${z_A,z_B,z_C}$ with nine shared initial-noise samples, yielding 27 utterances organized into nine matched triplets. Corresponding grid positions across the three latent conditions in Fig.~\ref{fig:cross_noise_reuse} use the same $x_0$, allowing latent-induced acoustic tendencies to be compared under matched initial noise.

Acoustic measurements show consistent latent-dependent tendencies across the nine matched noise samples. Utterance-level median $F_0$ follows the same ordering $z_A < z_B < z_C$, while mean frame energy consistently follows $z_B > z_C > z_A$.
The duration patterns provide complementary evidence:
$z_A$ consistently favors a longer UW1 in ``student'' and a
shorter AE1 in ``am'', whereas $z_B$ exhibits the opposite
preference, with neither pronounced pattern under $z_C$.
Because total generation length is fixed, these differences
reflect redistribution of within-utterance timing rather than
changes in overall sequence length.
Together, the reproducible pitch, energy, and timing
tendencies across matched $x_0$ samples provide evidence that the velocity model actively uses $z$ as a
trajectory-level realization condition whose effect persists
across initial-noise trajectories.

\section{Conclusion}
\label{sec}

We presented ReaFlow-TTS, a realization-conditioned
flow-matching framework that uses an utterance-level
stochastic latent to explicitly account for
realization-dependent variation in velocity prediction.
Holding the latent fixed throughout generation provides a
consistent realization condition across the flow trajectory,
while semantic calibration of the realization space enables
graded VAD manipulation.
Experiments show that realization conditioning improves
synthesis quality under full acoustic masking and induces
reproducible acoustic tendencies across initial-noise
trajectories, while learned VAD directions support graded manipulation across generation contexts with only modest changes in naturalness.
Together, the reproducible pitch, energy, and timing tendencies across nine matched initial-noise samples provide behavioral evidence that $z$ serves as a reusable realization condition across initial-noise trajectories.


\bibliographystyle{IEEEbib}
\bibliography{refs2}

\end{document}